\documentstyle[12pt]{article}
\input epsfig.sty      
\begin{document}


\title { {\tt \it \small Proceedings of International Conference on Magnetism 2000, 
to appear in J. Magn. Magn. Mater. } \\
{ \ } \\
Phase-coherence transition in granular superconductors with $\pi$ junctions}

\author{Enzo Granato \\
Laborat\'orio Associado de Sensores e Materiais, \\
Instituto Nacional de Pesquisas Espaciais, \\
12201-190 S\~ao  Jos\'e dos Campos, SP Brasil }

\date{}
\maketitle

\begin{abstract}                       
We study the three-dimensional XY-spin glass as a model for the resistive behavior of 
granular superconductors containing  a random distribution of $\pi$ junctions, as in 
high-$T_c$ superconducting materials with d-wave symmetry. The $\pi$ junctions leads to 
quenched in circulating currents (chiralities) and to a  chiral-glass state at low 
temperatures, even in the absence of an external magnetic field.   
Dynamical simulations in the phase representation are used to determine the 
nonlinear current-voltage characteristics as a function of temperature. 
Based on  dynamic scaling analysis, we find a phase-coherence transition at finite 
temperature below which the linear resistivity should vanish and determine the corresponding 
critical exponents. The results suggest that the phase and chiralities may order simultaneously 
for decreasing  temperatures into a superconducting chiral-glass state. 
\end{abstract}


\medskip






\newpage

Granular superconductors, including the high $T_c$ materials, 
are often described by arrays of superconducting grains coupled together
by  Josephson junctions. In a system of  conventional junctions ($s$-wave pairing)
the phases of the superconducting order parameter of neighboring grains  
tend to be locked with zero phase shift and a phase-coherence transition 
occurs for decreasing temperature into a state  
with long-range phase coherence and vanishing linear resistivity \cite{fisher}. 
On the other hand, there is increasing evidence for $d$-wave pairing symmetry 
in high $T_c$ materials with the remarkable consequence of the appearance of 
"$\pi$" junctions (negative coupling) characterized by a phase 
shift \cite{sigrist} of  $\pi$. 
This leads to frustration effects in granular samples, even in zero external 
magnetic field, since a closed loop containing an odd number of  $\pi$ junctions 
gives rise to an spontaneous circulating current (orbital magnetic moment) and can 
be a possible explanation of the  paramagnetic Meissner 
effect \cite{sigrist,kusmat,kawamura}. There is a close connection between this system and 
the XY-spin glass model \cite{kawamura} where the two-component spins, 
$s=(\cos(\theta),\sin(\theta))$,  correspond to superconducting grains and the 
random ferro or antiferromagnetic interactions to the Josephson couplings.
A chiral order parameter can be defined in the XY-spin glass model measuring the
direction of circulating current (vortex) in the loops for the superconducting array. 
Based on this analogy and numerical results for the XY-spin glass in 
three dimensions \cite{kawamura}, it has been argued that the equilibrium low-temperature
state is a chiral glass but with no long-range phase coherence and therefore 
not a true superconductor. However, recent results for XY-spin glass suggest that 
a spin glass (phase coherence) transition is possible \cite{grempel} in addition
to the  chiral glass transition. These results however rely on simulations for 
the ground state defect energy and spin-glass order parameter which are inaccessible
experimentally. Experiments often measure transport properties and it would be of interest
to study directly the current-voltage behavior from dynamical simulations and examine 
the issue of the finite-temperature resistive transition.  

To study the current-voltage characteristics we assume a resistively shunted 
Josephson-junction  model for the current flow between grains represented 
by the phases $\theta_i$ on a cubic lattice and  simulate the nonequilibrium behavior 
using the  Langevin equations \cite{eg}
\begin{equation}
C_o \frac{d^2 \theta_i}{dt^2} + \frac{1}{R_o}\sum_j \frac{d(\theta_i-\theta_j)}{dt} =
-J_o \sum_j \sin(\theta_i-\theta_j-A_{ij})+\sum_j \eta_{ij} 
\end{equation}
where $R_o$ is a shunt resistance, $C_o$ is the capacitance to the ground, 
$J_o>0$ is the Josephson coupling, $\eta_{ij}$ represents 
uncorrelated thermal noise with $<\eta_{ij}(t)^2>=2k_B T/R_o$ to ensure thermal 
equilibrium, and  $A_{ij}=0$ or $\pi $, represents the phase
shift across the junction, with equal probability of $0$ and $\pi$ corresponding to 
the standard XY spin glass \cite{kawamura,grempel,eg,wengel}. 
We use units where $\hbar/2e=1$, $R_o=1$, $I_o=1$ 
and set  $J_oR_o^2C_o=0.5$, corresponding to the overdamped regime. 
The equations were integrated numerically \cite{eg} for a system of linear size $L$
and the voltage $V$ (electric field $E/L$)  computed as
a function of the driving current $I$ (current density $J=I/L^2$) averaging 
over $5-10$ different realizations of the $A_{ij}$  distribution. 

The nonlinear resistivity $E/J$ as a function of temperature $T$ 
is shown in Fig. 1a, for $L=12$. The behavior is consistent with a linear resistivity 
$\rho_L=\lim_{J\rightarrow 0} E/J$ which is finite above a critical temperature
$T_c$. At lower temperatures, it appear to extrapolate to zero, indicating  
a superconducting transition with $T_c$ in the range $0.3 - 0.5$. 
This is confirmed by a scaling analysis of the nonlinear resistivity which
assumes the existence of a continuous superconducting transition at a finite temperature
\cite{fisher}. Near the transition, measurable quantities scale with the diverging correlation 
length $\xi\propto |T-T_c|^{-\nu}$ and relaxation time $\tau \propto \xi^z$, 
where $\nu$ and $z$ are the correlation-length and  dynamical critical exponents, 
respectively.  The nonlinear resistivity should then satisfy the scaling form \cite{fisher}
$ T E \xi^{z-d+2}/J= g_\pm(J\xi^{d-1}/T) $
where $d$ is the system dimension and $g(x)$ is a scaling function. The $+$ and $-$ 
signs correspond to $T>T_c$ and $T<T_c$, respectively. A scaling plot according to this
equation can be used to verify the scaling arguments 
and the assumption of an equilibrium transition at $J=0$ and also provide an estimate 
of the critical temperature and critical exponents. This is shown in Fig. 1b, 
obtained by adjusting 
the unknown parameters so that the best data collapse is obtained. We estimate
$T_c=0.41(3)$, $z=4.6(4)$ and $\nu = 1.2(4)$. The value of  $\nu$ 
agrees  with the result obtained from calculations of $\rho_L$ using the 
fluctuation-dissipation relation (zero current bias) in the vortex 
representation \cite{wengel}. However, $z$ is significantly larger 
but this could be a result of  different dynamics. 
We also note that $T_c$ and the critical exponents agree with the estimates for the 
gauge-glass model \cite{fisher} suggesting a common universality glass, although in 
the latter case $A_{ij}$ has a continuous distribution in the interval $[0,2\pi]$. 
Similar agreement has been found in the vortex representation \cite{wengel}. 
Since, in the phase model, 
the superconducting transition should correspond to the critical temperature 
where long-range  phase coherence sets in \cite{eg}, our results suggest a 
phase coherence transition at finite temperatures. 
This is in contrast with other numerical results indicating a phase coherence transition 
only  at zero temperature \cite{kawamura}. However, it is consistent with more
recent calculations  showing evidence of a  lower-critical dimension 
below or equal $3$ for the   XY-spin glass model \cite{grempel} and therefore a 
phase-coherence transition at finite temperature is possible. 
On the other hand, evidence of a chiral-glass transition has been found 
at $T_c\sim 0.32$ from Monte Carlo simulations \cite{kawamura}, 
below our estimate for the phase-coherence transition, which could suggest an 
intermediate phase. However, this 
requires phase coherence in presence of uncorrelated chiral (vortex) disorder which is 
unlikely. Thus, the finite-temperature transition found in the present  calculations 
should correspond to a single transition in the system where chirality and phase coherence
order simultaneously into a low temperature superconducting chiral glass phase. Although
our present conclusion is based on the scaling analysis of the nonlinear current-voltage 
characteristics which is a nonequilibrium property of the system, the good agreement with 
the previous results in the vortex representation  obtained from calculations of 
the (equilibrium) linear resistivity \cite{wengel}, suggests that this transition 
corresponds to the underlying equilibrium behavior. Additional calculations for the linear
resistivity will help to settle this interesting issue.

\medskip
This work was supported by FAPESP (grant: 99/02532).

\newpage


\bigskip
\begin{figure}
\centering\epsfig{file=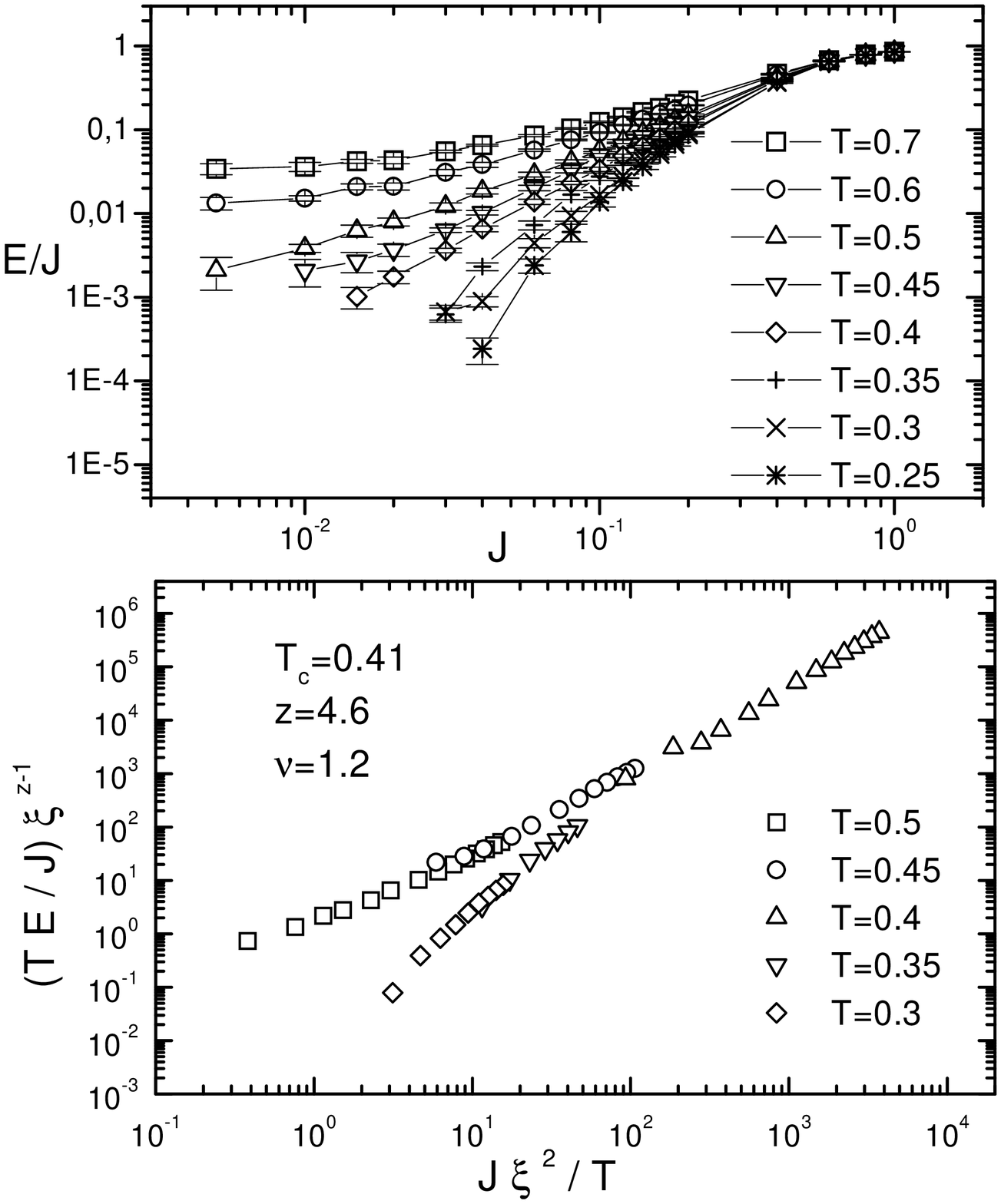,bbllx=1cm,bblly=0.5cm,bburx=20cm,bbury=26cm,width=8.5cm}
\caption{
Fig. 1. (a) Nonlinear resistivity $E/J$ as a function of temperature. (b) Scaling plot
near the transition obtained by adjusting $T_c$, $z$ and $\nu$ 
where $\xi\propto|T/T_c-1|^{-\nu}$. }
\end{figure}

\end{document}